# Central Kurdish Text-to-Speech Synthesis with Novel End-to-End Transformer Training

Hawraz A. Ahmad [1]  and Tarik A. Rashid [2,*]

[1] Department of Software and Informatics Engineering, Salahaddin University-Erbil, Erbil 44001, Iraq; hawraz.dizayi@su.edu.krd
[2] Department of Computer Science and Engineering, University of Kurdistan Hawler, Erbil 44001, Iraq
* Correspondence: tarik.ahmed@ukh.edu.krd

**Abstract:** Recent advancements in text-to-speech (TTS) models have aimed to streamline the two-stage process into a single-stage training approach. However, many single-stage models still lag behind in audio quality, particularly when handling Kurdish text and speech. There is a critical need to enhance text-to-speech conversion for the Kurdish language, particularly for the Sorani dialect, which has been relatively neglected and is underrepresented in recent text-to-speech advancements. This study introduces an end-to-end TTS model for efficiently generating high-quality Kurdish audio. The proposed method leverages a variational autoencoder (VAE) that is pre-trained for audio waveform reconstruction and is augmented by adversarial training. This involves aligning the prior distribution established by the pre-trained encoder with the posterior distribution of the text encoder within latent variables. Additionally, a stochastic duration predictor is incorporated to imbue synthesized Kurdish speech with diverse rhythms. By aligning latent distributions and integrating the stochastic duration predictor, the proposed method facilitates the real-time generation of natural Kurdish speech audio, offering flexibility in pitches and rhythms. Empirical evaluation via the mean opinion score (MOS) on a custom dataset confirms the superior performance of our approach (MOS of 3.94) compared with that of a one-stage system and other two-staged systems as assessed through a subjective human evaluation.

## 1. Introduction

In recent years, text-to-speech (TTS) systems have undergone significant advancements, which were largely driven by the adoption of deep learning (DL) techniques. These systems transform written text into natural-sounding speech through a multi-stage process. However, traditional approaches often necessitate trade-offs between the quality of the synthesized speech, the speed of generation, and the complexity of the training process.

One prevalent approach leverages a two-stage architecture [1,2]. In the initial stage, the system generates intermediate representations, such as linguistic features [2] or mel-spectrograms—a representation capturing the frequency content of an audio signal over time [1] from the preprocessed text. To generate mel-spectrograms from the preprocessed text, the system typically utilizes techniques like text-to-speech (TTS) models that convert the text into spectrogram images. These spectrograms represent how the frequency content of the audio signal varies over time, providing a structured format for subsequent stages of audio synthesis. The second stage then translates these representations into raw audio waveforms [2,3]. While this method has yielded realistic speech, it suffers from limitations. Training these models often requires sequential training or fine-tuning, hindering efficiency. Additionally, their reliance on predefined intermediate features restricts the utilization of potentially beneficial learned representations, limiting the system's ability to further improve performance.

To address these shortcomings, alternative approaches have been explored, such as non-autoregressive methods [4,5] and generative adversarial networks (GANs) [6]. Non-autoregressive models aim to overcome the slow generation speed inherent to autoregressive systems such as Tacotron 2 [1] and Transformer TTS [7] by eliminating the sequential nature of the process. These models can synthesize speech significantly faster, making them more suitable for real-time applications. On the other hand, GAN-based methods have shown promise in generating high-quality waveforms, potentially surpassing the quality achieved by two-stage approaches [8,9]. Recent efforts have focused on developing efficient end-to-end training methods for TTS models [10,11]. These methods aim to bypass the two-stage pipeline entirely and directly convert text into speech. While these approaches offer potential performance improvements by leveraging learned representations throughout the entire process, they often fall short of the quality achieved by two-stage systems, highlighting the ongoing challenge of balancing efficiency and quality in TTS.

The Sorani Kurdish dialect is predominantly spoken by Kurdish communities in Iraq and Iran, representing a significant linguistic and cultural identity within these regions. Despite its importance, Sorani Kurdish has faced challenges in modern technological advancements, particularly in the realm of TTS conversion. TTS technology plays a crucial role in enhancing accessibility and inclusivity for languages and dialects worldwide. However, the development of TTS systems for Sorani Kurdish has been relatively limited compared to more widely spoken languages. This discrepancy poses a barrier to the full participation of Sorani Kurdish speakers in digital communication, education, and accessibility tools that rely on TTS technology.

This study presents a parallel end-to-end TTS method to address the limitations of both traditional two-stage architectures and recent end-to-end methods. The proposed model is named Kurdish TTS (KTTS), as it will be developed to generate more natural-sounding Central Kurdish audio than that generated by current two-stage models. The goal of this study is to achieve high-quality speech synthesis while maintaining efficiency and simplified training protocols. Our approach utilizes a variational autoencoder (VAE) that is pre-trained for audio waveform reconstruction. This involves aligning the prior distribution established by the pre-trained encoder with the posterior distribution of the text encoder within latent variables. To augment the expressive capabilities of our method and enable the synthesis of high-fidelity speech waveforms, we employ adversarial training [6] in the waveform reconstruction. By aligning latent distributions and integrating the stochastic duration predictor, our method facilitates the real-time generation of natural Kurdish audio speech. The proposed model is trained directly to maximize the log-likelihood of speech, and this is coupled with the alignment process.

One of the crucial steps in developing a TTS system is the creation of a high-quality speech corpus. Developing models that capture the prosodic patterns of Kurdish is essential for creating natural-sounding synthesized speech. Other contributions of this research work can be summarized as follows:

- A novel end-to-end method for Kurdish text-to-speech conversion based on a VAE framework is introduced. The proposed VAE effectively maps input waveforms to a latent space representation and reconstructs them.
- A robust training procedure is developed to align the latent variables of the text encoder with those of the pre-trained waveform encoder of the VAE. This involves ensuring that the prior distribution established by the pre-trained encoder matches with the posterior distribution of the text encoder within latent variables.
- The proposed KTTS directly regenerates waveforms from input text by bypassing the intermediate stages required to create mel-spectrograms or linguistic features.
- A new dataset comprising aligned pairs of Central Kurdish text sequences and corresponding audio recordings is curated. This dataset serves as a valuable resource for advancing research in Kurdish text-to-speech synthesis.

The rest of this article is organized as follows: Section 2 offers a review of the TTS literature, Section 3 explains the methodology for KTTS, and Section 4 details the experimental setup, including the dataset and training parameters. In Section 5, we present the results and a discussion, and finally, Section 6 provides conclusions based on our findings.

**2. TTS Literature Review**

This section provides an overview of existing TTS systems, categorizing them into one-stage and two-stage approaches. It also briefly discusses the existing literature on Kurdish TTS systems. Transformers, initially introduced by Vaswani et al. in 2017 [12], have revolutionized natural language processing by capturing long-range dependencies among input tokens, which is particularly beneficial for tasks like text-to-speech (TTS) synthesis. In recent years, their application to TTS development has yielded significant improvements in the naturalness and intelligibility of synthesized speech. This advance underscores the versatility and power of transformers in handling complex sequential data, highlightingtheir potential in other domains as well. In the following two subsections, we summarize state-of-the-art TTS systems, which are mostly based on transformer architectures.

*2.1. One-Stage Systems*

One-stage text-to-speech (TTS) systems streamline the process of converting text into synthetic speech by employing a direct transformation model. These systems leverage end-to-end neural network architectures, such as sequence-to-sequence models with attention mechanisms, to map raw text directly to acoustic waveforms. In variational inference with adversarial learning for end-to-end text-to-speech (VITS) [13], a duration predictor is introduced to improve the rhythm of the generated utterances. VITS was extended to allow the generation of diverse utterances for multi-language speakers using your-TTS. Although these models allow sampling from the input tokens, the quality of these generated utterances is still inferior to that obtained with single-speaker systems [14].

In ref. [4], the authors introduced FastSpeech, a non-autoregressive version of transformer TTS [7]. They used the original model as a teacher and extracted the character durations from it. To generate all output frames, they trained a student model using a convolutional duration prediction.

In 2021, Ren et al. introduced Fast Speech 2 [10], a non-autoregressive version of transformer TTS. The researchers used external durations to improve the training process and reduce the development costs. This approach assumes that the alignment model used for the language is of high quality.

Recent advancements include models like VITS 2 [15], which combines variational inference with normalizing flows and adversarial learning to directly generate high-fidelity speech waveforms from text. In ref. [16], the authors proposed a framework for building controllable TTS systems that can generate speech with specific attributes. It combines a sequence-to-sequence TTS model with a conditional variational autoencoder (CVAE) to learn disentangled representations of speech attributes. The system enables flexible and controllable speech synthesis. The integration of large language models, such as LLaMA, into TTS systems, has been shown to enhance semantic understanding and generate more expressive speech [17], highlighting the potential for semantic-aware TTS systems to further improve synthesis quality.

Our proposed model lies in the category of one-stage systems, where no intermediate stages are needed to create mel-spectrograms and then to convert mel-spectrograms into waveforms, as our system regenerates waveforms directly from input text.

*2.2. Two-Stage Systems*

Two-stage TTS systems introduce an intermediate step between text processing and waveform generation, typically predicting a mel-spectrogram before synthesizing the final speech output. This approach separates the linguistic and acoustic modeling stages, allowing for more fine-grained control and potentially higher-quality synthetic speech.

The SC-GlowTTS system [18] is a flow-based multi-speaker text recognition sys- tem that takes the predicted parameters of an external speaker embedding into account. SNAC [19], on the other hand, utilizes a coupling layer to explicitly normalize the input.

The basic Glow-TTS [20] architecture consists of a flow-based determinate duration predictor, a transformer-equipped encoder, and a flow-dependent decoder. The transformer-based encoder produces a linear approximation of the prior distribution mean by translating the input tokens' phonetic embedding into a representation with an 80-dimensional structure. The z-sampling method can also be utilized to express the distribution's z-sampled value:

$$z = \mu + T\epsilon, \quad (1)$$

In training, the duration predictor only predicts the mean μ and the temperature T, while at inference time, it chooses a value of T that is usually smaller than 1. A latent representation of the distribution is then sampled from the prior data to generate a mel-spectrogram.

In 2017, Vaswani et al. introduced the concept of transformer TTS [12]. In 2019, Li et al. [7] tested the effectiveness of this technology by developing an algorithm that can predict the mel-spectrogram for English phonemes. The evaluation of the transformer TTS system showed that it was very promising, but it was not feasible to use it in a production setting because the auto-regressive approach was time-consuming. The evaluation of the mean opinion score (MOS) of a phonemicized dataset using a non-autoregressive model was not significantly different from the results when using the transformer TTS system. The authors also used a pitch prediction module with FastPitch [21] to complement the duration predictor in their work from the Tactron2 model that they introduced [1]. The authors claimed that the quality of their results was similar when using durations and phonemes from a Montreal forced alignment (MFA) model [22].

Diff-TTS [23] uses a diffusion probabilistic model to first generate mel-spectrograms, which are then converted to speech using advanced vocoder models like HiFi-GAN [9]. More recently, models such as WaveGrad 2 [24] and EfficientTTS 2 [25] focus on optimizing the two-stage process for faster and more efficient synthesis without compromising on quality. In 2023, the MelStyleTTS [26] proposed a style transfer technique for mel-spectrograms, allowing for greater expressiveness in synthetic speech. Two-stage systems have been shown to produce more natural and expressive speech compared to their one-stage counterparts, although they may introduce additional latency and complexity in the synthesis pipeline.

## 3. Related Work for the Kurdish Language

After conducting an extensive review of the existing literature on the Kurdish language, it is apparent that the majority of previous research has focused on utilizing and adapting existing models rather than developing novel approaches. We have summarized some of the works in Table 1, which shows some initial works in Kurdish and some Kurdish TTS approaches.

**Table 1.** Summary of the main points of the Kurdish literature review.

| No. | Reference | Year | Method | Dataset | Result |
|---|---|---|---|---|---|
| 1 | [27] | 2009 | Concatenative (Allophone, Syllable, and Diphone) | Kurdish Language | Allophone MOS 2.45 Syllable MOS 3.02 Diphone MOS 3.51 |
| 2 | [28] | 2009 | Concatenative (Allophone, Syllable, and Diphone) | Kurdish Language | Best quality score 3.5 Best DRT 97% |
| 3 | [29] | 2009 | Concatenative (Allophone) | Kurdish Language (2100 words) | Best quality score 2.4 |

**Table 1.** *Cont.*

| No. | Reference | Year | Method | Dataset | Result |
|---|---|---|---|---|---|
| 4 | [30] | 2011 | Concatenative (Diphone) | Kurdish Language (2100 words) | Best quality score 55% |
| 5 | [31] | 2020 | Tacotron 2-Transfer Learning | Nawar Halabi's Arabic Dataset | (3 h) MOS 4.21 |
| 6 | [32] | 2022 | Tacotron 2 | Persian dataset | (21 h) MOS 3.01–3.97 |

## 4. Methodology

This section explains the proposed method and its architecture. As illustrated in Figure 1, our approach for Kurdish text-to-speech conversion comprises three key procedures: VAE for waveform reconstruction (Figure 1A), training (Figure 1B), and inference (Figure 1A). Detailed descriptions of the components and blocks employed within our framework will be explained in this section.

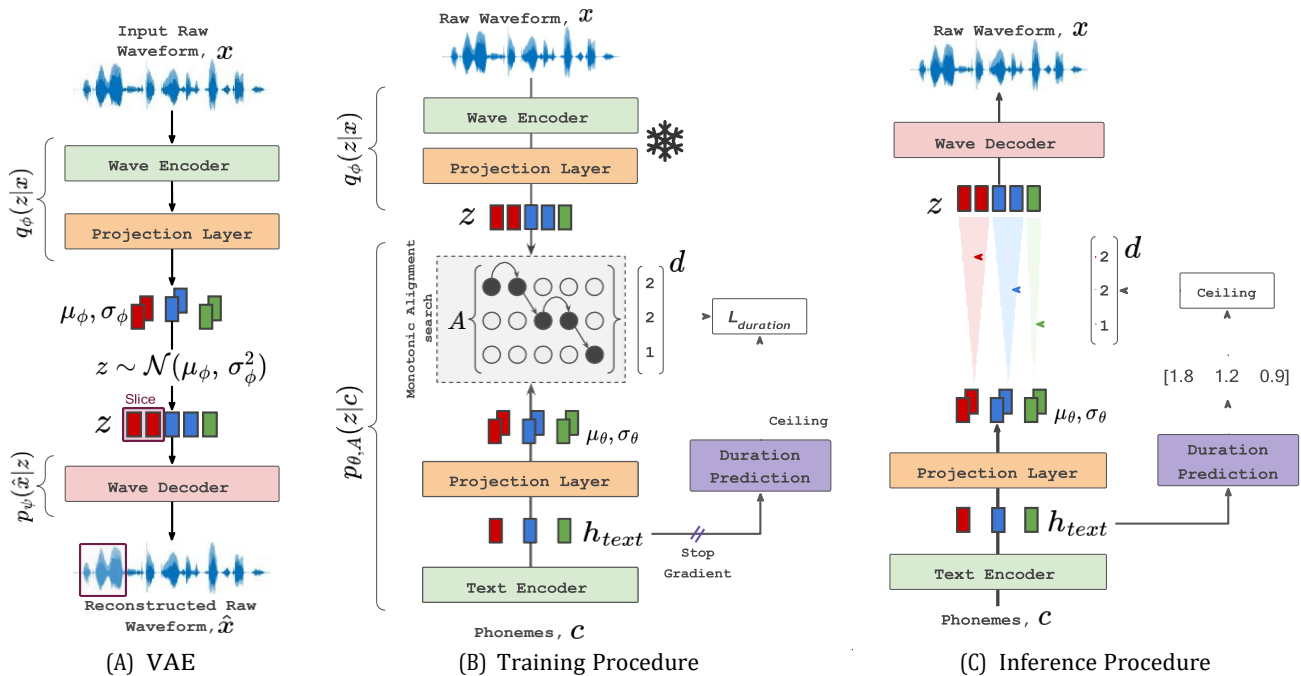

**Figure 1.** (**A**) Initially, a VAE is pre-trained using speech-to-speech data. During this phase, the VAE focuses on reconstructing the input speech waveform. (**B**) Training Procedure: This is the alignment phase where the pre-trained wave encoder of the VAE is utilized to ensure that the text encoder produces a distribution of latent variables identical to that generated by the wave encoder. (**C**) Inference Procedure: In this phase, the text encoder is trained to generate distributions that the wave decoder of the VAE can interpret and convert into speech waveforms.

### 4.1. Variational Autoencoder

To effectively pre-train a VAE for waveform reconstruction, several key components and formulas are essential. The VAE framework aims to learn a probabilistic mapping from an input waveform $x$ to a latent space representation $z$ and subsequently reconstructs the input as $\hat{x}$. This process involves two primary objectives: maximizing the likelihood of generating the input data given the latent variables and enforcing the learned latent space to follow a prior distribution.

The wave encoder $q_\phi(z|x)$, an approximate posterior distribution that is parameterized by $\phi$, maps the input waveform $x$ to a latent space representation $z$, where $z \sim \mathcal{N}(\mu_\phi, \sigma_\phi^2)$, with $\mu_\phi$ and $\sigma_\phi^2$ representing the mean and variance of the latent space distribution, respectively. Subsequently, the wave decoder $p_\psi(\hat{x}|z)$ parameterized by $\psi$ generates the reconstructed output $\hat{x}$ conditioned on the sampled latent variable $z$.

For training our VAE, the total loss $L(\psi, \phi)$ is the sum of two terms: the reconstruction loss $L_{rec}$ and Kullback–Leibler (*KL*) divergence.

$$L(\psi, \phi) = - E_{z \sim q_\phi(z|x)}[\log p_\psi(x|z)] + D_{KL}[q_\phi(z|x)||p_\psi(z)] \quad (2)$$

$L_{rec}$ is expressed by the negative log-likelihood $-\log p_\psi(x|z)$ capturing the probabilistic aspect of reconstruction, ensuring that the generated output closely resembles the input data distribution. For $L_{rec}$, we simply use the mean square error (MSE) between the input $x$ and the reconstructed output $\hat{x}$. Second, the *KL* divergence term enforces a regularization constraint, guiding the latent space towards a predefined prior distribution $p_\psi(z)$, which is a Gaussian distribution of $N(0, 1)$. Now, the target is to find the optimal $\psi$ and $\phi$ such that

$$\psi^*, \phi^* = \arg\max_{\psi, \phi} L(\psi, \phi) \quad (3)$$

### 4.2. Training Procedure

During the training, we take the pre-trained wave encoder $q_\phi(z|x)$ from the previous step with its parameters $\phi$ frozen to generate the latent representation $z$ of the target waveform $x$.

Our ultimate goal is to model the conditional distribution of the waveform data $p_{\theta,A,\psi}(x|c)$ by transforming a conditional prior distribution $p_{\theta,A}(z|c)$ through the pre-trained wave decoder $p_\psi(x|z)$, where $c$ represents the input text sequence (see Figure 1C). We parameterize the prior distribution with the parameters $\theta$ of the text encoder and an alignment function $A$, which is discussed in Section 4.3.

To achieve this, we need to minimize the distance between the posterior distribution of the pre-trained wave encoder $p_\phi(z|x)$ and the prior distribution $p_{\theta,A}(z|c)$. Once again, we employ KL divergence to force the latent space $z$ to conform to $p_\phi(z|x)$. The KL divergence is, then,

$$D_{KL} = \log q_\phi(z|x) - \log p_{\theta,A}(z|c) \quad (4)$$

where

$$z \sim q_\phi(z|x) = N(z; \mu_\phi(x), \sigma_\phi(x)) \quad (5)$$

The prior distribution's statistics, denoted as $\mu_\theta$ and $\sigma_\theta$, are computed using the text encoder, which can transform the text condition $c = c_{1:T_{text}}$ into the corresponding statistics, $\mu = \mu_{1:T_{text}}$ and $\sigma = \sigma_{1:T_{text}}$, with $T_{text}$ representing the length of the input text.

### 4.3. Alignment Prediction

The alignment function $A$ denotes the mapping from the index of the latent representation of waveform $z$ to the corresponding index of statistics from the text encoder, $A(j) = i$, whenever $z_j$ follows a normal distribution $N(z_j; \mu_i, \sigma_i)$. We presume that $A$ maintains both monotonicity and subjectivity to prevent skipping or repeating the input text. Subsequently, the prior distribution can be articulated as follows:

$$\log p_{\theta,A}(z|c) = \sum_{j}^{T_{waveform}} \log N(z_j; \mu_{A(j)}, \sigma_{A(j)}), \quad (6)$$

where $T_{waveform}$ is the length of the input waveform.

Similar to [20], we employ a monotonic alignment search to find the parameters $\theta$ and the alignment $A$ that maximize the log-likelihood of waveform data, as shown in Equation (7).

$$\max_{\theta, A} L(\theta, A) = \max_{\theta, A} p_{\theta, A, \psi}(x|c) \quad (7)$$

Throughout the training process, we keep the parameters of the pre-trained VAE $\phi$ and $\psi$ frozen. Consequently, our objective is to find the optimal alignment function $A^*$, after which we update $\theta$ using gradient descent:

$$\begin{aligned} A^* &= \arg\max_A \log p_{\theta, A, \psi}(x|c) \\ &= \arg\max_A \sum_j^{T_{waveform}} \log \mathsf{N}(z_j; \mu_{A(j)}, \sigma_{A(j)}) \end{aligned} \quad (8)$$

### 4.4. Duration Prediction

Given the absence of ground-truth labels for the alignment, it becomes necessary to estimate the alignment at every training iteration. The duration of each input token $d_i$ can be computed by summing the columns within each row of the estimated alignment, as shown in Equation (9). This duration calculation serves as our ground truth for training a deterministic duration predictor, $f_{duration}$.

$$d_i = \sum_j^{T_{waveform}} \mathbb{1}_{A^*(j)=i}, i = 1, 2, \ldots T_{text} \quad (9)$$

During the training procedure, we train $f_{duration}$ to re-predict the duration computed in Equation (9) from the optimal alignment $A^*$. This duration prediction also helps predict $A^*$ during the inference process. We train $f_{duration}$ with the MSE loss, as outlined in Equation (10), by integrating it on top of the text encoder (see Figure 1). In order to prevent interference with the maximum likelihood objective, we employ the stop gradient technique on the input of the duration predictor during the backward pass [33].

$$\mathsf{L}_{duration} = MSE(D, d), \quad (10)$$

where

$$D = \lceil f_{duration}(SG(h)) \rceil \quad (11)$$

where $SG$ denotes the stop gradient operator, and $h_{text}$ is the hidden representation of the text encoder.

### 4.5. Inference Procedure

Throughout the inference process, which is illustrated in Figure 1C, the statistical parameters $\mu_\theta$ and $\sigma_\theta$ of the prior distribution, along with $A^*$, are obtained by the text encoder and duration predictor. Then, a latent variable is sampled from the prior distribution $z \sim \mathsf{N}(\mu_\theta, \sigma_\theta^2)$, and concurrently, a waveform $\hat{x}$ is synthesized by transforming the sampled $z$ using the pre-trained wave decoder. Instead of feeding the entire latent representation $z$, we segment $z$ into slices with a size of 32, each corresponding to a brief audio clip. The pre-trained wave decoder sequentially receives the slices and up-samples (transforms) them to the corresponding audio clips.

### 4.6. Model Architecture
#### 4.6.1. Text Encoder

To handle Central Kurdish text, our initial step involves converting text sequences into International Phonetic Alphabet (IPA) sequences through the utilization of open-source software [34]. Additionally, we incorporate several custom-defined phonemes to accommodate the distinct characters present in Central Kurdish, as outlined in Appendix A.2. Then, the text encoder converts the phoneme embedding sequence into the hidden phoneme representation $h_{text}$. We follow the encoder structure of the transformer [12], as shown in Figure 2, with some slight modifications. We remove the positional encoding and add learnable positional encoding. We build the text encoder with eight blocks of transformer encoders, each with eight multi-head self-attention modules. The dimension of phoneme embeddings and the hidden size of the self-attention (hidden representations) are set to 256 following the recommendation by FASTSPEECH [10].

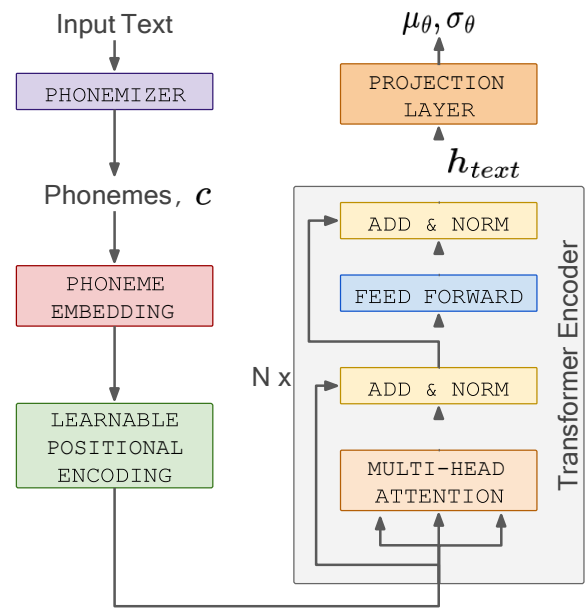

**Figure 2.** The text encoder featuring a modified transformer encoder with learnable positional encoding.

The positional encoder depicted in Figure 3 employs a grouped 1D convolution comprising 64 filters with a kernel size of 3 to generate a relative positional vector from the latent features. This vector is subsequently combined with the embedding of phonemes (tokens) to encode their positions relative to each other. We append a linear projection layer on top of the transformer encoder to predict the statistics of the prior distribution, $\mu_\theta$ and $\sigma_\theta$, from the hidden representation $h_{text}$.

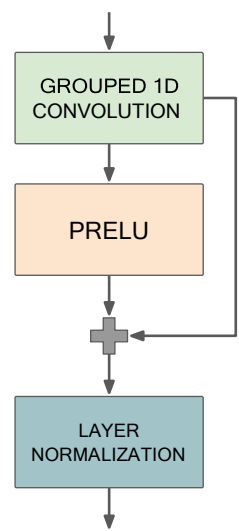

**Figure 3.** The positional encoder uses a 64-filter grouped 1D convolution to generate relative positional vectors.

### 4.6.2. Wave Encoder

To build our wave encoder, we utilize a transformer encoder structure identical to that employed in the text encoder, as depicted in Figure 4. This choice aids the text encoder network in converging more swiftly when the KL divergence is applied for the difference between the two distributions.

In designing the feature encoding block, we adopt a similar structure to that outlined in ref. [35] with slight changes, as depicted in Figure 5. To enhance the processing efficiency, we opted for a configuration of five 1D convolutional blocks instead of the original seven, achieving comparable results according to empirical validation. Additionally, we substitute the GEUL activation layers with PRELU. The receptive field of the feature encoder spans a total context of 2200 samples, corresponding to 100 ms at the 22 kHz input sample rate. Consequently, this feature encoder extracts features from the raw waveform and tokenizes it, with each token representing a 100 ms segment. These tokens are subsequently processed by the transformer encoder.

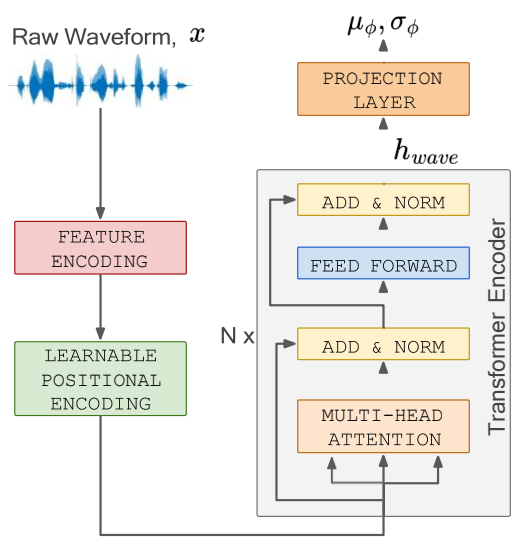

**Figure 4.** The wave encoder architecture utilizes a transformer structure akin to that of the text encoder, enhancing convergence during the application of KL divergence.

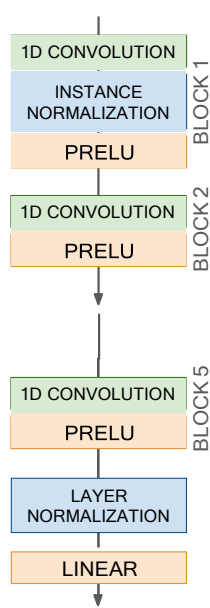

**Figure 5.** The feature encoder with a 2200-sample receptive field tokenizing 100 ms segments of raw waveforms for transformer processing.

#### 4.6.3. Wave Decoder

Our wave decoder architecture is modeled after WaveNet [2], as illustrated in Figure 6. It comprises a transposed 1D convolution with a filter size of 64, along with 30 dilated residual convolution blocks. The skip channel size and kernel size of the 1D convolution are configured to 64 and 3, respectively. The wave decoder receives a sliced hidden representation $z$ with a channel size of 256 generated by the wave or text encoder, corresponding to a brief audio clip, as its input. It then utilizes transposed 1D convolution to upsample the slice, aligning it with the length of the corresponding audio clip.

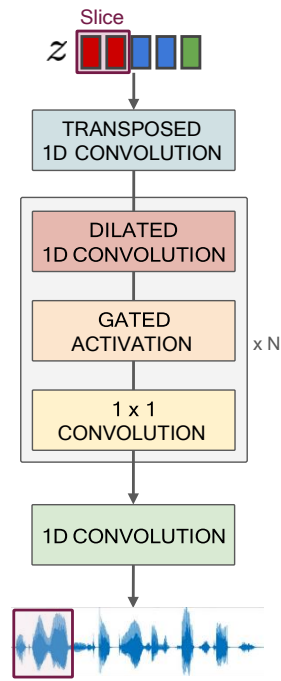

**Figure 6.** The wave decoder architecture inspired by WaveNet [2], featuring transposed 1D convolution, and dilated residual blocks with adversarial training for enhanced waveform generation.

Similar to prior works [2,13,36], we incorporate adversarial training into the wave decoder. The discriminator $D$ in the adversarial training adopts the same structure and configurations as those of Parallel WaveGAN [36]. $D$ distinguishes between the waveform $\hat{x}$ generated by the wave decoder and the ground-truth waveform $x$. We optimize the wave decoder by incorporating the multi-resolution short-time Fourier transform (STFT) loss in conjunction with the discriminator loss from the least squares generative adversarial network (LSGAN), aligning with the methodology of Parallel WaveGAN [36].

#### 4.6.4. Duration Predictor

$f_{duration}$ predicts the distribution of phoneme durations from the hidden representation $h_{text}$. To build its architecture, we stack two residual blocks, as shown in Figure 7. Each of these blocks consists of a convolutional layer containing 256 filters, each with a kernel size of 3, alongside a PRELU activation function and layer normalization followed by an FC. PRELU is chosen for its ability to learn negative slope values, mitigating the issue of deadneurons associated with RELU. Additionally, the inclusion of residual connections serves to mitigate vanishing gradients, thereby enhancing performance and reducing overfittingby encouraging feature reuse.

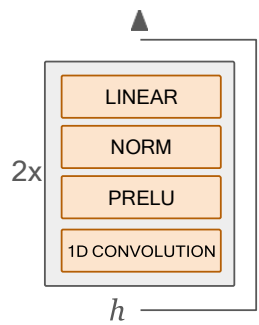

**Figure 7.** The architecture of the duration prediction model $f_{duration}$.

## 5. Experiments

This section details the dataset creation, categorization, and recording process, followed by an overview of the corpus statistics and technical specifications. The training approach, including the dataset partitioning and pre-training of the VAE, is outlined, along with the GPU usage and optimization techniques, providing a concise overview of our experimental setup.

### 5.1. Dataset

Text-to-speech systems depend on the availability of a corpus containing pairs of speech and corresponding text. This study explores voice data of the Central Kurdish dialect for TTS systems. We started by creating an audio- and text-pairing dataset featuring a male individual who spoke in Central Kurdish. The recording process was carried out by a male dubber in a recording studio. The 6078 sentences that we collected from the text corpus were categorized into 12 categories, including sports, science, literature, health, and everything else. Training sentences were then created using the collected information, resulting in 4255 (70%) sentences. The validation set contained 608 (10%) random sentences. The testing set contained 1215 (20%) sentences that were randomly selected from the overall dataset. The sentences were then improved through various web sources. Table 2 illustrates the subjects and the number of sentences. The process of recording speech files ended after 30 days. The dataset can be accessed through the following link [37].

**Table 2.** Statistics of the sentences of the speech corpus.

| Topics | No. of Sentences |
|---|---|
| News | 888 |
| Sport | 631 |
| Health | 463 |
| Interview | 1240 |
| Science | 65 |
| Religion | 24 |
| Economic | 275 |
| General information | 224 |
| Politics | 66 |
| Education and literature | 1399 |
| Article | 420 |
| Social | 383 |
| Total | 6078 |

Some features of the recorded files can be summarized as follows: (a) 6078 WAV files and over 13.63 h of recorded speech were captured; (b) the output of the files was recorded at a rate of 22,050 kHz; (c) the quantization process was carried out using 16 bits of signed data; (d) the stored speech audio files were in the format known as PCM, and a mono channel was utilized to record the audio streams; (e) the shortest audio file length was 0.502 s; (f) the longest audio file length was 16.781 s; (g) the mean audio file length was 8.076 s.

The audio files are stored in wave format, while the text sentences are saved in an Excel file. The audio files are organized in a single folder. The audio file's name includes the extension names, while the transcript is the text of the speech referenced to the audio file with an ID which is the name of the audio file. The dataset was prepared to comply with the Gaussian distribution to be more effective in training models avoiding bias in record length. A statistical figure of the dataset has been created to show more clarity on the number of audio records of similar length recordings as depicted in Figure 8.

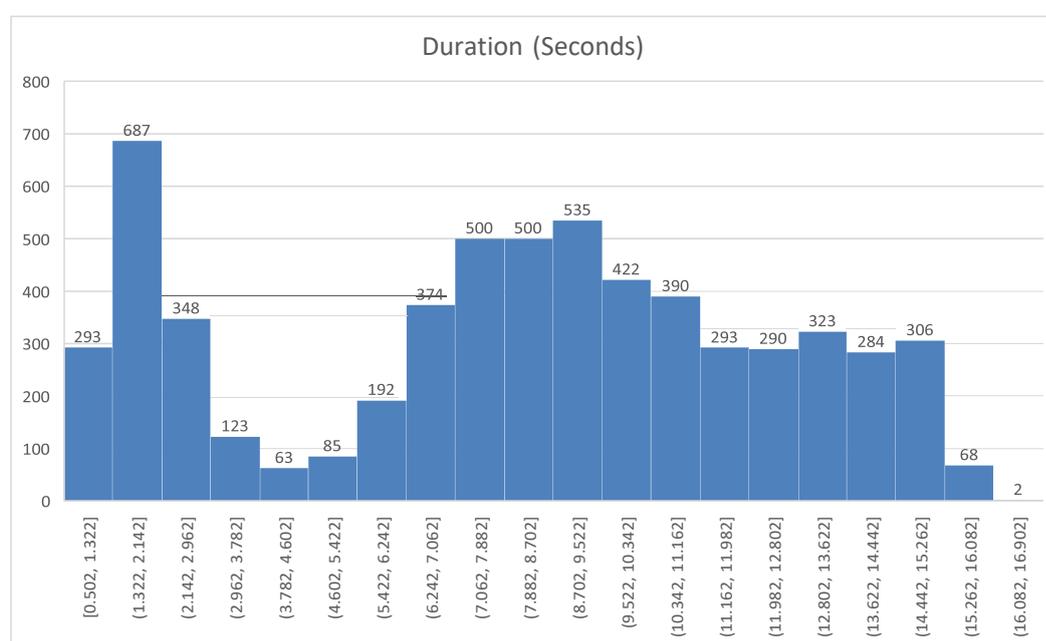

**Figure 8.** Histogram of Kurdish sentences in the dataset.

*5.2. Training*

We first split the dataset into the following three subsets: 70% for training, 10% for validation, and 20% for testing. Before starting the training procedure to align the wave and text encoders, we pre-trained the VAE using only the audio waveforms. The VAE takes an input audio $x$ and attempts to reconstruct $\hat{x}$ after compressing $x$ into $z$. To reduce the training time, memory usage, and complexity, we fed a randomly selected sliced hidden representation $z$ with a window size of 32 to the wave decoder. To compute the STFT and LSGAN losses, we extracted the corresponding audio segments from the ground-truth raw waveforms as training targets. We followed the Parallel WaveGAN [36] for the details of the adversarial training.

After the VAE had converged, we utilized the pre-trained wave encoder to initiate the training procedure. The aim was to align the latent distribution of the text encoder with that of the pre-trained wave encoder so that it could later be recognized by the pre-trained wave decoder.

The training of both the VAE and the alignment of the wave and text encoders was conducted on two RTX A5000 GPUs manufactured by NVIDIA sourced from Denver, Colorado in United States of America. The VAE was trained using a batch size of 18 waveforms per GPU. The optimization was performed by utilizing the Adam optimizer [38] with the parameters set to $\beta_1 = 0.9$, $\beta_2 = 0.98$, and $\epsilon = 10^{-9}$. The learning rate decay was scheduled by a factor of 0.9991/8 per epoch, starting from an initial value of $1 \times 10^{-3}$. It took 430 K steps for training until convergence.

The training for the alignment of the wave and text encoders was executed with a batch size of 12 sentences per GPU by utilizing the AdamW optimizer [39] with the parameters $\beta_1 = 0.8$, $\beta_2 = 0.98$, and a weight decay of $\lambda = 0.01$. The learning rate decay followed a schedule of 0.9991/8 factors per epoch, starting from an initial learning rate of $2 \times 10^{-4}$. The training process reached convergence after 820 K steps.

## 6. Results and Discussion

In this section, we evaluate the performance of KTTS in terms of audio quality and inference speed.

### 6.1. Audio Quality

We evaluated the generated synthetic audio files in the test set to obtain the MOS to measure the audio quality. We kept the sentence content consistent among the different models so as to exclude other interference factors and avoid biases by only examining the audio quality. Each audio was listened to by at least 54 evaluators who were all native speakers of Central Kurdish. We compared the MOSs of the audio samples generated by our KTTS model with those of other well-known TTS models, which included (a) the GT (the ground-truth audio), (b) Tacotron 2 [1] (Mel + multi-band MelGAN [40]), (c) VITS [13], a conditional variational autoencoder with adversarial learning for end-to-end text-to-speech conversion, and (d) Glow TTS [20], a generative flow for text-to-speech conversion via monotonic alignment search (Mel + multi-band MelGAN). For each model, only ratings greater than one were considered, while those equal to or below this threshold were excluded from the analysis. The results are shown in Table 3. It can be seen that our KTTS outperformed the mentioned one-stage TTS system, and it reached the quality of the two-stage TTS systems.

**Table 3.** Comparison of the evaluated MOSs with 95% confidence intervals on the Gigant dataset.

| Model | MOS (CI) |
|---|---|
| Ground Truth | 4.75 (±0.10) |
| Tacotron 2 (Mel + multi-band MelGAN) | 3.85 (±0.10) |
| Glow TTS (Mel + multi-band MelGAN) | 3.94 (±0.15) |
| VITS | 3.70 (±0.66) |
| **KTTS** | **3.94 (±0.16)** |

### 6.2. Inference Speed

This section compares the inference speed of KTTS with that of both two-stage and one-stage systems. The comparison was conducted on a server with an "AMD Ryzen threadripper pro 3955wx" CPU with 16 cores, 256 GB of memory, and one NVIDIA RTX A5000 GPU with 24 GB of memory. Table 4 shows that the proposed model sped up the inference process by 8.32x compared with that of the one-stage VITS system [13]. Regarding the two-stage systems, the proposed model sped up the inference process by 47.49x with respect to Glow TTS [20] and by 53.73x with respect to Tacotron 2 [1], as the two-stage systems needed more processing time because two parallel models were included in their inference processes.

**Table 4.** A comparison of the inference speed with 95% confidence intervals.

| Model | Latency (s) | Speedup (%) |
|---|---|---|
| One-Stage Systems | | |
| VITS | 0.560 ± 0.068 | / |
| KTTS | 0.517 ± 0.093 | 8.32 |
| Two-Stage Systems | | |
| Tacotron 2 (Mel + multi-band MelGAN) | 0.795 ± 0.011 | / |
| KTTS | 0.517 ± 0.093 | 53.73 |
| Glow TTS (Mel + multi-band MelGAN) | 0.763 ± 0.044 | / |
| KTTS | 0.517 ± 0.093 | 47.49 |

A real-time factor (RTF) comparison was conducted in order to evaluate the model's efficiency in synthesizing speech. Table 5 shows that the proposed model outperformed the other one-stage and two-stage models in real time.

**Table 5.** Comparison of RTF with 95% confidence intervals.

| Model | RTF (CI) |
|---|---|
| Tacotron 2 (Mel + multi-band MelGAN) | 0.099 ± 0.006 |
| Glow TTS (Mel + multi-band MelGAN) | 0.095 ± 0.011 |
| VITS | 0.0701 ± 0.006 |
| **KTTS** | **0.065 ± 0.004** |

Since our approach relies on VAEs, it is important to acknowledge some inherent limitations of this method. VAEs, though promising in various applications, including TTS, face certain challenges. Being an unsupervised learning system, VAEs lack precise control over the speech features they generate. A key drawback of VAEs is the need to balance regularization and reconstruction accuracy, which can sometimes lead to distorted speech outputs. Additionally, the alignment between synthesized speech and input text is not explicitly defined in VAEs. Moreover, VAEs require meticulous tuning of several hyperparameters, such as the dimensionality of the latent space, the choice of a prior distribution, and the weighting of loss and reconstruction terms. These factors necessitate careful consideration to optimize performance and mitigate potential issues.

The proposed approach may yield suboptimal results under certain conditions, such as a lack of sufficient training data, complex phonetic variability, and real-time constraints. These conditions can involve complex computations and latent space sampling during inference, which may be computationally expensive.

## 7. Conclusions

This study introduced KTTS, an efficient end-to-end text-to-speech model tailored for generating high-quality Kurdish audio. By leveraging a pre-trained VAE for audio waveform reconstruction and integrating adversarial training techniques, we enhanced the expressiveness while ensuring high-fidelity speech synthesis. Our methodology effectively bridges the text-to-speech gap by aligning prior and posterior distributions within latent variables. An empirical evaluation on a custom dataset demonstrated KTTS's superior performance, which was comparable to the ground truth and was validated through subjective human evaluation. This represents a significant advancement in TTS technology, offering efficiency, quality, and flexibility for the synthesis of Kurdish text and speech. Future work may focus on enhancing the model further and expanding its applicability to other Kurdish dialects, speech styles, and multi-speaker models. Another area for future work involves integrating the two training phases of the VAE and KTTS into a unified procedure, enabling seamless integration without the requirement to pre-train the VAE network's encoder and decoder separately.

**Author Contributions:** H.A.A.: Data curation, original draft preparation, visualization, conceptualization, methodology, software, validation, resources, writing, reviewing and editing. T.A.R.: supervision, reviewing, and editing. All authors have read and agreed to the published version of the manuscript.

**Funding:** This research received no external funding.

**Data Availability Statement:** The data is accessible via the Mendeley Data repository at the following URL: https://data.mendeley.com/datasets/zhnvwsd7hs/1 (accessed on 18 June 2024).

**Acknowledgments:** We would like to acknowledge Jegr Nadhim for his invaluable assistance in preparing the server with two NVIDIA GPUs for our development process.

**Conflicts of Interest:** The authors declare no conflicts of interest.

## Abbreviations

The following abbreviations are used in this manuscript:

| | |
|---|---|
| TTS | Text to speech |
| MOS | Mean opinion score |
| VAE | Variational autoencoder |
| GANs | Generative adversarial networks |
| KTTS | Kurdish text to speech |

## Appendix A

*Appendix A.1. Kurdish Language*

This section covers the three major Kurdish dialects and provides an extensive elaboration on the writing system of the Central Kurdish dialect. It also delves into the various pronunciation points and grammatical structure of the language. Most Kurds living in Iraq and Iran use the Central Kurdish dialect. It became an official language of Iraq in 2006. The Northern Kurdish dialect, on the other hand, is commonly spoken in areas such as Syria, Turkey, and Northwestern Iran, among others. Southern Kurdish is also known as Zazaki or Hawrami, and it can be heard in areas such as Ilam and Kermanshah in Iran. The main types of Kurdish are Central and Northern Kurdish, which are, respectively, written in the Latin and Arabic alphabets.

Despite the number of speakers, the number of resources for the Central Kurdish dialect is still more than sufficient. The system for writing the language was first established during the 1920s, and it has undergone numerous changes. Table A1 provides a brief overview of the phonological features of the phonetic alphabet system.

The writing system of Central Kurdish is similar to that of a phonemic system. The assigned letter of the language is referred to as a "phoneme". Some exceptions are made. For example, the letter "ی" is pronounced as palatal approximants "/j/", and the "/i/" as it is referred to as a "vowel". The letter "و" is also called a bilabial approximant, and it is pronounced as a a vowel: "/u/" or "/ᴜ/". The letter "و" can be written in a repetitive way, " وو ", which is pronounced as "/u/", a long vowel. Although it is one phoneme, it is not considered as a separate character in the keyboard layout released by the Department of Information Technology of the Regional Government of Kurdistan in 2014.

Although the Central Kurdish system is similar to that of Persian and Arabic writing, there are differences. Table A2 illustrates the differences between the three alphabets, as they can be used to distinguish among different languages. In Arabic and Persian, there are usually problems with the homograph and Kasre, but in Kurdish writing, these issues do not occur, unlike in other languages, such as Arabic, Persian, and English. Due to the mapping between the written and spoken terms, these issues are not as common.

**Table A1.** Central Kurdish phoneme list for the e-speak phonemizer.

| No | Feature | | IPA | Phoneme | Letter (isolated form) | Example |
|---|---|---|---|---|---|---|
| 1 | Voiced Stop | | b | b | ب | ابوک |
| 2 | | | d | d | د | کستان |
| 3 | | | d͡ʒ | dʒ | ج | مانه |
| 4 | | | g | g | گ | گۆڕه |
| 5 | Voiced Fricative | | v | v | ڤ | هەڤژین |
| 6 | | | z | z | ز | حەز |
| 7 | | | ʒ | ʒ | ژ | ژن |
| 8 | | | ɣ | gh | غ | ردان |
| 9 | | | ʕ | β | ع | یب |
| 10 | | | t | t | ت | اتریک |
| 11 | unvoiced Stop | | t͡ʃ | tʃ | چ | م |
| 12 | | | k | k | ک | دار |
| 13 | unvoiced Fricative | | h | h | ھ | Mە |
| 14 | | | ʃ | ʃ | ش | ن |
| 15 | | | s | s | س | ر |
| 16 | Vibrant Flap | | ɾ | ɾ | ر | وڕ |
| 17 | Vibrant Trill | | r | r | ڕ | پڕ |
| 18 | Lateral | | l | l | ل | کل |
| 19 | | | l | ɫ | ڵ | واڵ |
| 20 | Nasal | | m | m | م | م |
| 21 | | | n | n | ن | ونوسین |
| 22 | Approximant | | j | j | ی | کی |
| 23 | | | w | w | و | وت |
| 24 | Vowels | | | | | |
| 25 | Front | High | i | j | ی | هەنوی |
| 26 | Central | Low | ä | a | ا | ابران |
| 27 | Front | Mid-low | ɛ | e | ئ | هەئوئ |
| 28 | Back | Mid | o | o | ۆ | ۆ |
| 29 | Central-back | Mid-high | ʊ | w | و | وۆ |
| 30 | Back | High | u | ww | و | ولوت |
| 31 | Front | Low | a | ʌ | ە | پش |
| 32 | Central-front | Mid-high | I | | | ێ |

Speech synthesis is a multidisciplinary field that has a wide range of problems. Among these is the issue with the management of unfamiliar words, as well as the prosody of proper nouns and foreign ones. In addition, synthesizers commonly encounter issues with the wave concatenation technique due to the varying effects of contextual and thematic factors [41].

It can be very challenging to create text due to the various language-related aspects involved. For instance, every non-standard term should have a phonetic equivalent. In addition, full words should be made from numbers and letters [42].

In Central Kurdish, some of the issues that can be encountered are hidden short-vowel characters. For instance, the words (ئەم , ئەز , سە) include hidden short vowels between consonant letters [43].

In contrast to other languages, such as Persian, English, and Arabic, the problems encountered by speech synthesizer users in Kurdish are not as common. Speech synthesizer users have to recognize the prosodic elements of written text, including the intonation, stress, and length. The features of continuous speech are influenced by the personality and emotions of the artist. Unfortunately, there is a lack of knowledge regarding the prosodic elements in written texts, which causes many of them to be frequently modified as speech is synthesized.

**Table A2.** Letters of Kurdish in comparison with Persian and Arabic letters [43].

| Language | Letters |
|---|---|
| Kurdish only | /a/ە /o/ۆ /e/ێ /ll/ڵ /rr/ڕ /v/ڤ |
| Kurdish and Persian | /zh/ژ /g/گ /ch/چ /p/پ |
| Kurdish, Persian and Arabic | /xe/غ /ah/ع /sh/ش /s/س /z/ز /r/ر /d/د /kh/خ /he/ح /je/ج /t/ت /b/ب /aa/ا /eh/ء /y/ی /h/ھ /w/و /n/ن /m/م /l/ل /k/ک /q/ق /f/ف |
| Persian and Arabic | (shaddah)◌ّ /o/◌ُ /e/◌ِ /a/◌َ /t/ط /z/ظ /dh/ذ /d/ض /s/ص /th/ث |

*Appendix A.2. Phonemization*

The phonemization process is the act of breaking words down into their phonemes. Phonemes are the smallest unit of speech sound, and they differentiate one word from another in a particular language. In our work, we used the e-speak phonemizer [34] to break the words down into their corresponding phonemes. We created a list of Central Kurdish letter phonemes that did not exist in the original repository. The phoneme list is illustrated in Table A1. As can be easily seen, we assigned some different phonemes from those in the IPA to comply with the e-speak package, as well as to overcome some pronunciation errors that we encountered during testing.